\documentclass[runningheads]{llncs}
\PassOptionsToPackage{numbers, compress}{natbib}
\usepackage{algorithm}
\usepackage{algorithmic}
\usepackage[hyphens]{url}
\usepackage{amsmath,amssymb,amsfonts}

\usepackage{graphicx}
\begin{document}
\title{TFCNs: A CNN-Transformer Hybrid Network for Medical Image Segmentation}
\titlerunning{TFCNs}

\author{Zihan Li\inst{\#1} \and
Dihan Li\inst{\#1} \and
Cangbai Xu\inst{1} \and Weice Wang\inst{1} \and Qingqi Hong\inst{*1,4}\orcidID{0000-0002-9996-6870} \and Qingde Li\inst{2}\orcidID{0000-0001-5998-7565} \and Jie Tian\inst{3}\orcidID{0000-0003-0498-0432}}

\renewcommand{\thefootnote}{\fnsymbol{footnote}}
\footnotetext{\# means equal contribution, * means corresponding author}

\authorrunning{Zihan Li et al.}
\institute {Xiamen University, Xiamen, 361005, China
\email{\{zihanli,dihanli,cangbaixu,wangweice\}@stu.xmu.edu.cn}
\email{hongqq@xmu.edu.cn}
\and {University of Hull, Hull, HU6 7RX, UK
\email{Q.li@hull.ac.uk}}
 \and Chinese Academy of Sciences, Beijing, 100190, China
\email{tian@ieee.org}  \and State Key Laboratory of Virtual Reality Technology and Systems, Beihang University, China }

\maketitle              % typeset the header of the contribution
\begin{abstract}
Medical image segmentation is one of the most fundamental tasks concerning medical information analysis. Various solutions have been proposed so far, including many deep learning-based techniques, such as U-Net, FC-DenseNet, etc. However, high-precision medical image segmentation remains a highly challenging task due to the existence of inherent magnification and distortion in medical images as well as the presence of lesions with similar density to normal tissues. In this paper, we propose TFCNs (Transformers for Fully Convolutional denseNets) to tackle the problem by introducing ResLinear-Transformer (RL-Transformer) and Convolutional Linear Attention Block (CLAB) to FC-DenseNet. TFCNs is not only able to utilize more latent information from the CT images for feature extraction, but also can capture and disseminate semantic features and filter non-semantic features more effectively through the CLAB module. Our experimental results show that TFCNs can achieve state-of-the-art performance with dice scores of 83.72\% on the Synapse dataset. In addition, we evaluate the robustness of TFCNs for lesion area effects on the COVID-19 public datasets. The Python code will be made publicly available on \url{https://github.com/HUANGLIZI/TFCNs}.
\end{abstract}
\section{Introduction}
\label{sec1}
Medical image segmentation plays a critical role in clinical diagnosis and assisting doctors to evaluate patient’s reactions to treatment. Various algorithms based on convolutional neural networks (CNNs) \cite{lecun1998gradient} have been applied to image segmentation. And with a U-shaped network design, U-Net \cite{ronneberger2015u} has achieved great success in various medical imaging applications. Following this technical route, many algorithms have been developed for medical image and volume segmentation, such as U-Net++ \cite{zhou2018unet++}. In order to solve the degradation problem, He et al. proposed ResNets \cite{he2016identity}, which aims to simplify very deep networks by introducing a residual block that sums two input signals. Then a new CNN architecture called DenseNets  \cite{iandola2014densenet} has been developed by the composition of dense blocks and pooling operations. In FC-DenseNet \cite{zhang2018automatic}, the up-sampling path was introduced to restore the input resolution. Recently, inspired by the great success of Transformers in the field of natural language processing (NLP) \cite{chowdhury2003natural}, researchers have tried to introduce Transformers into the field of computer vision \cite{li2022lvit,zhou2021deepvit}. Vision transformer (ViT) \cite{dosovitskiy2020image} has been proposed to achieve object detection tasks.

\begin{figure}[H]
\centerline{\includegraphics[width=0.8\textwidth]{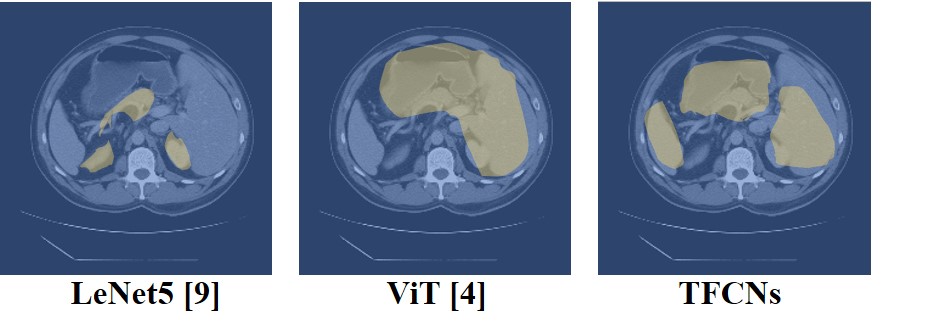}}
\vspace{-10pt}
\caption{Class activation maps in LeNet5 \cite{lecun1998gradient}, ViT \cite{dosovitskiy2020image} and TFCNs by using the CAM method \cite{zhou2016learning}. In which we set a fixed value as the activation intensity threshold.}
\label{fig1}
\end{figure}

Currently, there are two problems: 1). As shown in Fig.\ref{fig1}, since the convolution operation collects information by layer, which leads to too much focus on local feature information. In the field of medical image segmentation, the lack of global information often leads to false category of segmentation. Therefore, a visual transformer was introduced, which can reflect complex spatial transformations and long-distance feature dependencies, which are regarded as global representations. Currently, although Chen et al. proposed Transunet \cite{chen2021transunet} to solve problems such as lack of high-level details. However, we found that the direct feeding of CNN-style features into the transformer for recoding tends to bring limited improvement. 2). In U-shape networks, the skip connection between encoder and decoder is crucial. However, semantic-independent features tend to be fed to the decoder with direct transmission, which will interfere with image segmentation. Our main motivation is how to preserve image features more completely.
\vspace{-10pt}
\vspace{-20pt}
To tackle the problem 1), we utilize DenseBlocks to facilitate the propagation of feature information to the transformer part while adding a residual structure to the MLP to further effectively preserve the global representation information. To tackle the problem 2), we decide to fuse spatial and channel attention on the original skip connection to transmit information more efficiently. Our main contributions are as follows:
\begin{enumerate}
\item A new deep neural network framework (TFCNs) is proposed, to the best of our knowledge, which is the first network to introduce Transformers into FC-DenseNet and improve the internal structure of Transformers.
\item A general attention module CLAB (Convolutional Linear Attention Block) is proposed to improve segmentation performance, which includes two types of attentions: (a) attention over the spatial extent of image and (b) attention over the CNN-style feature channels.
\end{enumerate}

\section{Related Work}
\label{sec2}
In the field of semantic segmentation, FCNs \cite{long2015fully} innovatively proposed a model structure in the form of encoder-decoder. And to solve the problem of information loss in the encoding process, it utilized the form of residual connection to combine the encoding process. In addition to the semantic segmentation of real objects, more and more attention has also been paid to medical image segmentation. Based on FCNs,  U-Net \cite{ronneberger2015u} was proposed and applied to medical image segmentation. This model makes use of a mutually symmetrical encoding-decoding design. Another example was FC-DenseNet \cite{jegou2017one}, where they extended the work in DenseNets \cite{iandola2014densenet} by introducing the DenseBlock in the process of upsampling, which not only alleviated the problem of dimensional explosion in the deep encoder but also retained contextual information better. Some Transformer-based methods have also been proposed for semantic segmentation, object detection, and instance segmentation, such as SETR, DETR \cite{zheng2021rethinking,carion2020end}. Inspired by the previous breakthroughs, TransUNet \cite{chen2021transunet} embedded the Transformers in the down-sampling process to extract the information in the original image. More recently, a Gated Axial-Attention model was proposed in MedT \cite{valanarasu2021medical} to extend some existing attention-based schemes. There are also other variants to the Transformers such as the Swin Transformer \cite{liu2021swin}, which utilize a sliding window to limit self-attention calculations to non-overlapping partial windows.

\section{Method}
\label{sec3}

\begin{figure*}
\centerline{\includegraphics[width=\textwidth]{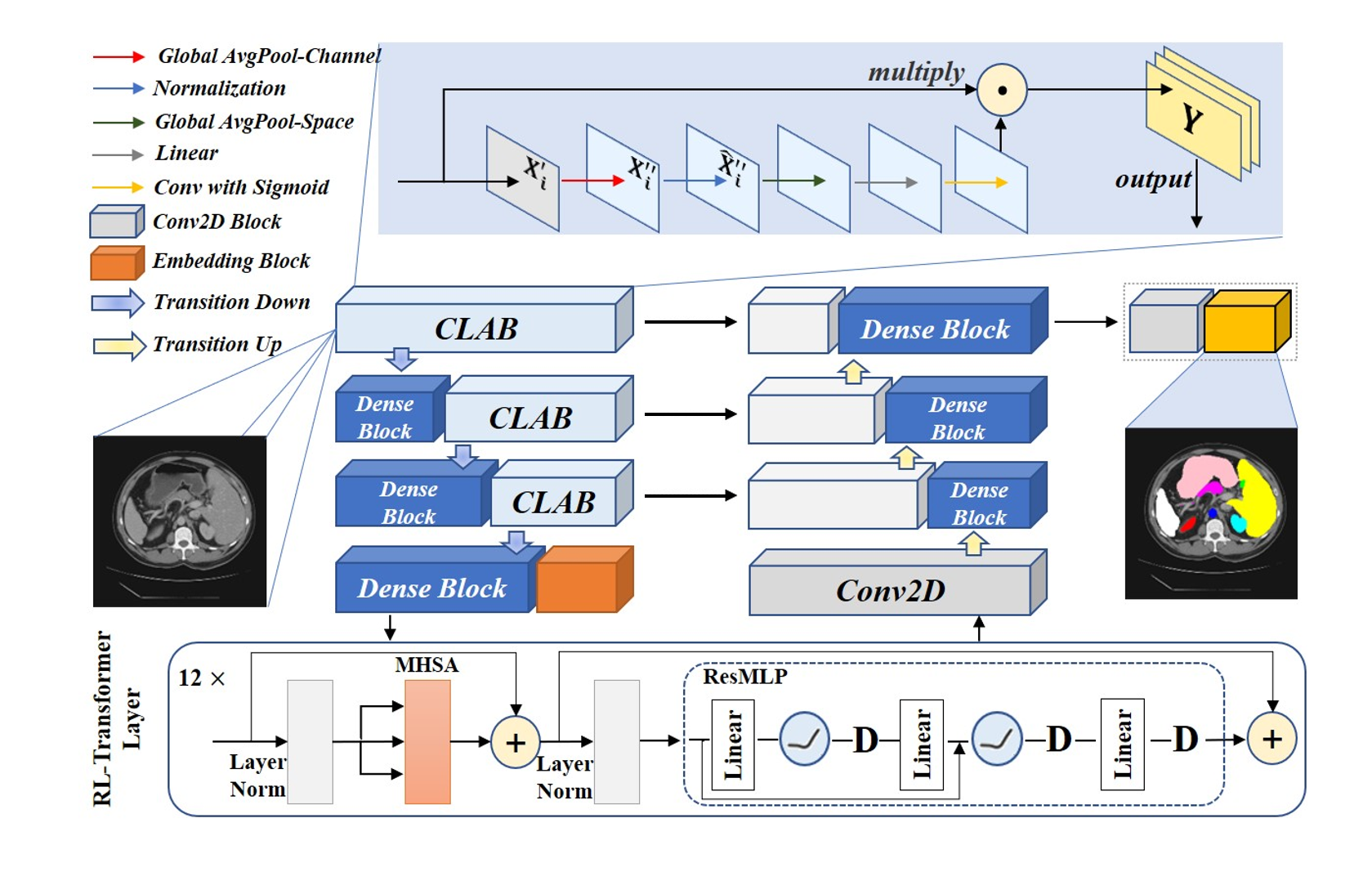}}
\caption{Overview of the proposed TFCNs. RL-Transformer module at the end of encoder gives access to a receptive field containing images and CLAB modules are dedicated to filtering non-semantic features by including spatial and channel attention.}
\label{fig2}
\end{figure*}

\subsection{Overall structure of TFCNs}
\label{sec3.1}
As described in the first section, the conventional U-shaped structured network lacks global contextual information to perform high-precision medical image segmentation. Given this, we propose TFCNs (Fig.\ref{fig2}), which takes FC-DenseNet \cite{zhang2018automatic} as the backbone network, with an RL-Transformer Layer being added to the encoder to enhance the segmentation capability of the network. In addition, CLAB (Convolutional Linear Attention Block) in the skip connection part is introduced to enhance the spatial recovery of the focused segmentation region. The CNN-Transformer hybrid model acts as an encoder and CLAB as the upper and lower connecting part between DenseBlock and Transition-Down, which helps to filter non-semantic features. Compared with TransUNet \cite{chen2021transunet}, TFCNs not only replaces the traditional convolutional layers with Dense Blocks, but also changes the feature encoding method. The details of each part of the structure will be described in the next two subsections. More specifically, the RL-Transformer (ResLinear-Transformer) is described in section \ref{sec3.2}, the CLAB (Convolutional Linear Attention Block) is described in section \ref{sec3.3}.

\subsection{RL-Transformer}
\label{sec3.2}
Referring to the ViT \cite{dosovitskiy2020image} implementation, we propose the ResLinear-Transformer (RL-Transformer) and apply it to the encoder of TFCNs. Of particular note is that the 2D patch $x_{P}^{i}$ is expanded linearly and its projection is mapped to the \emph{D} dimension using a trainable linear layer, as shown in Eqn.\ref{eq1}. The output of this projection is called Patch Embedding.
\begin{equation}
{z_0} = [x_{class};x_P^1E;x_P^2E; \cdots ;x_P^NE] + {E_{pos}}
\label{eq1}
\end{equation}
where $E$ is the patch embedding projection which is located before entering the RL-transform layer and ${E_{pos}}$ is the position embedding. RL-Transformer Layer consists of alternating ResMLP blocks and \emph{L} layers of Multi-Head Self-Attention (MHSA). The expressions are shown in Eqn.\ref{eq2} and Eqn.\ref{eq3}, where LN denotes Layer Norm.
\begin{equation}
{z'_\ell } = MHSA(LN({z_{\ell  - 1}})) + {z_{\ell  - 1}}
\label{eq2}
\end{equation}
\begin{equation}
{z_\ell } = ResMLP(LN({z'_\ell })) + {z'_\ell }
\label{eq3}
\end{equation}

The RL-Transformer encoder consists of alternating multi-headed self-attentive layers and ResMLP blocks. As shown in Figure \ref{fig2}, our proposed ResMLP consists of two GELU \cite{hendrycks2016gaussian} nonlinear layers, three Linear layers, and three dropout layers alternating with a residual connection to the source input before the second GELU \cite{hendrycks2016gaussian} layer. ResMLP can be expressed in Eqn.\ref{eq4} and Eqn.\ref{eq5} shown below:
\begin{equation}
{z''_\ell } = LN({z'_\ell })
\label{eq4}
\end{equation}
\begin{equation}
y = {z''_\ell } + L(\alpha GELU(L({z''_\ell })))
\label{eq5}
\end{equation}
where GELU represents the GELU\cite{hendrycks2016gaussian} nonlinear layer, \emph{L} represents the Linear layer, and $\alpha$ represents the associated weight parameters which is a learned parameter. Finally, the state of the sequence at the output of the RL-Transformer encoder is utilized as the image features.

\subsection{CLAB (Convolutional Linear Attention Block)}
\label{sec3.3}
Inspired by TTD (Test-Time Dropout) \cite{2018Aleatoric} and TTA (Test-Time Augmentation) \cite{2014Evaluation}, we propose \textbf{Convolutional Linear Attention Block} (CLAB). Nowadays TTD \cite{2018Aleatoric} can utilize dropout layers in the reasoning process and generate multiple predictions for each data instance. Compared with CUAB \cite{wang2021cuab}, we change the order of extracting channel and spatial attention, as indicated by the experimental results of CBAM \cite{woo2018cbam}. In addition, the normalization operation is utilized between the two attention operations to accelerate the training convergence process. Finally, inspired by the local field of view, the weak influence outside the local area is directly reduced to zero influence by using the linear layer.

As shown in Fig.\ref{fig2}, the global average pooling layer with the linear layer is added between the two convolutional layers. In CLAB, we first use $1 \times 1$ convolutional layers, where each convolutional layer has $K$ kernels to generate the sequence ${X'_i} = {\mathbb{R}^{H \times W \times K}},i \in \{ 1, \ldots ,N\} $. Then a global average pooling operation is performed on ${X'_i}$ in channel dimensions to obtain ${X''_i}$, and a normalization operation is performed on ${X''_i}$ to homogenize the data to obtain ${\hat X''_i} = \frac{{{{X'}_i} - {\mu _{\rm B}}}}{{\sqrt {\sigma _B^2 + \varepsilon } }}$ to accelerate the convergence and accuracy improvement of the training process, where ${\mu _B}$ denotes the mean of a batch, ${\sigma _B}$ denotes the standard deviation of the batch, and  $\varepsilon$ is a minimal positive value to ensure that the denominator is not zero. All ${\hat X''_i} = {\mathbb{R}^{H \times W}},i \in \{ 1, \ldots ,N\} $, are concatenated to form ${X_m} = {\mathbb{R}^{H \times W \times N}}$, which is then input sequentially into a submodule consisting of a global average pooling layer (with respect to dimensionality), a linear layer and a convolutional layer with sigmoid. The result is then multiplied with the source input to obtain the final output feature $Y$. We later perform ablation analysis for CLAB as well, and the experimental results show that CLAB has a significant effect on the model segmentation performance improvement.

\section{Experimental Results and Discussion}
\label{sec4}

\subsection{Implementation Details}
\label{sec4.1}
For all experiments, we perform a simple data augmentation, i.e., by performing random rotation or flipping operations. The optimizer chosen is the SGD optimizer, with a momentum of 0.9 and a weight decay of 1e-4. The learning rate selected for the experiment for our method is 0.005 and is set to 0.001 for other models. The learning rate is made to decay after 30,000 iterations. The batch size is set to 3 for Segtran \cite{zheng2021rethinking}, and 12 for all other models. The epoch is set to 150 on all datasets.

For Patch Size,it is worth noticing from Table \ref{table7} that the performance of the model is optimal when the patch size is set to 16. When the patch size is set to 8, the included area is too small, so some relatively large organs such as liver or kidney will be divided into many different patches for encoding, which splits some important information. This makes it difficult for the decoder to perform well, thereby reducing the performance of the model. When the patch size is set to 32, because the coding area is relatively large, it contains many interfering information that makes the model to misjudge. Although the CLAB module functions as a filter, the remaining redundant information is still greater than when the patch size is 16, so it also affects the judgments of the model.
\begin{table}[H]\centering
\caption{Ablation study on different patch sizes in transformer( Dice score\% and Hausdorff distance in mm and Jaccard score\%).}
\label{table7}
\begin{tabular}{c|c|c|c}
\hline
    Patch\_Size & Dice$\uparrow$           & Hd95$\downarrow$           & Jaccard$\uparrow$        \\ \hline
    8           & 78.79          & 38.11          & 65.95          \\
    16          & \textbf{83.72} & \textbf{17.26} & \textbf{72.78} \\
    32          & 80.00          & 24.41          & 67.84         \\ \hline
    \end{tabular}
\end{table}

Moreover, in our experiment, the combination of the Dice coefficient and cross-entropy is taken as the loss function for all methods. And the weight of these two components is 0.5.

\subsection{Comparison with other SOTA methods}
\label{sec4.2}

We conduct our experiments on Synapse\footnote{https://www.synapse.org/\#!Synapse:syn3193805/wiki/217789} dataset and COVID-19\footnote{https://aistudio.baidu.com/aistudio/datasetdetail/34221} dataset. The experimental results are analyzed by taking the Dice coefficient, Hausdorff distance, and Jaccard coefficient as evaluation factors. All the State-of-the-art methods are implemented using original paper without any deviations.

\noindent\textbf{Synapse Dataset}\par

As shown in Table \ref{table1}, these methods including Transformers, i.e. TransUNet \cite{chen2021transunet}, Segtran \cite{zheng2021rethinking}, and our TFCNs achieves better performance when compare with other methods. Especially, our method has a slight lead in terms of Dice coefficient and Jaccard coefficient when compared to TransUNet \cite{chen2021transunet} and Segtran \cite{zheng2021rethinking}. We believe this improvement is brought by the Dense Block we add, which is able to enhance the transmission of semantic information in the main pipeline.

\begin{table*}\centering
\caption{Performance comparison between our method and other state-of-the-art methods on Synapse dataset (Dice score\% and Hausdorff distance in mm and Jaccard score\%, and Dice score \% for each organ). Avg means average result of all testing cases and the Dice coefficient\% on each organ.}
\label{table1}
\resizebox{\columnwidth}{!}{
\begin{tabular}{c|c|c|c|c|c|c|c|c|c|c|c}
\hline
    Method      & Dice(avg)$\uparrow$      & Hd95(avg)$\downarrow$ & Jaccard(avg)$\uparrow$   & Aorta$\uparrow$          & Gallbladder$\uparrow$    & Kidney(L)$\uparrow$ & Kidney(R)$\uparrow$ & Liver$\uparrow$ & Pancreas$\uparrow$       & Spleen$\uparrow$         & Stomach$\uparrow$ \\ \hline
    U-Net \cite{ronneberger2015u}        & 81.81          & 26.81     & 71.21          & 86.99          & 67.31          & 88.64     & 82.81     & 94.15 & 60.99          & 90.50          & 83.10   \\
    Fc-DenseNet \cite{zhang2018automatic} & 81.62          & 21.83     & 70.62          & 86.68          & 66.31          & 88.14     & 82.07     & 95.26 & 61.73          & 92.11          & 80.68   \\
    AttU-Net \cite{oktay2018attention}  & 81.05          & 29.09     & 70.71          & 89.63          & 67.05          & 88.46     & 77.08     & 94.52 & 56.89          & 92.13          & 82.69   \\
    Res-UNet \cite{xiao2018weighted}     & 78.33          & 58.66     & 66.61          & 86.16          & 59.63          & 86.55     & 83.93     & 94.49 & 48.94          & 86.96          & 79.96   \\
    U-Net++ \cite{zhou2018unet++}      & 81.60          & 28.31     & 70.48          & 88.15          & 67.29          & 86.31     & 83.62     & 94.00 & 61.71          & 89.51          & 82.18   \\
    DDANet \cite{tomar2020ddanet}      & 79.60          & 21.29     & 67.87          & 83.74          & 64.93          & 88.93     & 83.83     & 94.99 & 51.99          & 90.68          & 77.69   \\
    TransUNet \cite{chen2021transunet}   & 82.33          & 19.88     & 71.18          & 88.50          & 62.96          & 91.23     & 90.03     & 94.90 & 59.90          & 90.53          & 80.59   \\
    Segtran \cite{zheng2021rethinking}     & 83.02          & 14.73     & 72.68          & 87.10          & 62.88          & 92.66     & 89.94     & 95.47 & 61.76          & 91.47          & 83.40   \\
    TFCNs       & \textbf{83.72} & 17.26     & \textbf{72.78} & \textbf{89.69} & \textbf{67.75} & 90.11     & 88.30     & 94.74 & \textbf{64.08} & \textbf{92.22} & 82.84 \\ \hline
    \end{tabular}}
\end{table*}

From Table \ref{table1}, it can be observed that our method achieves more accurate results than other approaches on some organs that are difficult to be segment, such as the Gallbladder and Pancreas. Since these tiny organs occupy a relatively small proportion in the original image, other approaches are easy to wrap other interfering information when extracting features from the areas containing these tiny organs. Conversely, through the utilization of the designed CLAB module, our model is able to pay more attention to these tiny organs themselves instead of those irrelevant areas. Meanwhile, for other organs, the results achieved by our method are also in the top 5.

\begin{figure}[H]
\centerline{\includegraphics[width=\columnwidth]{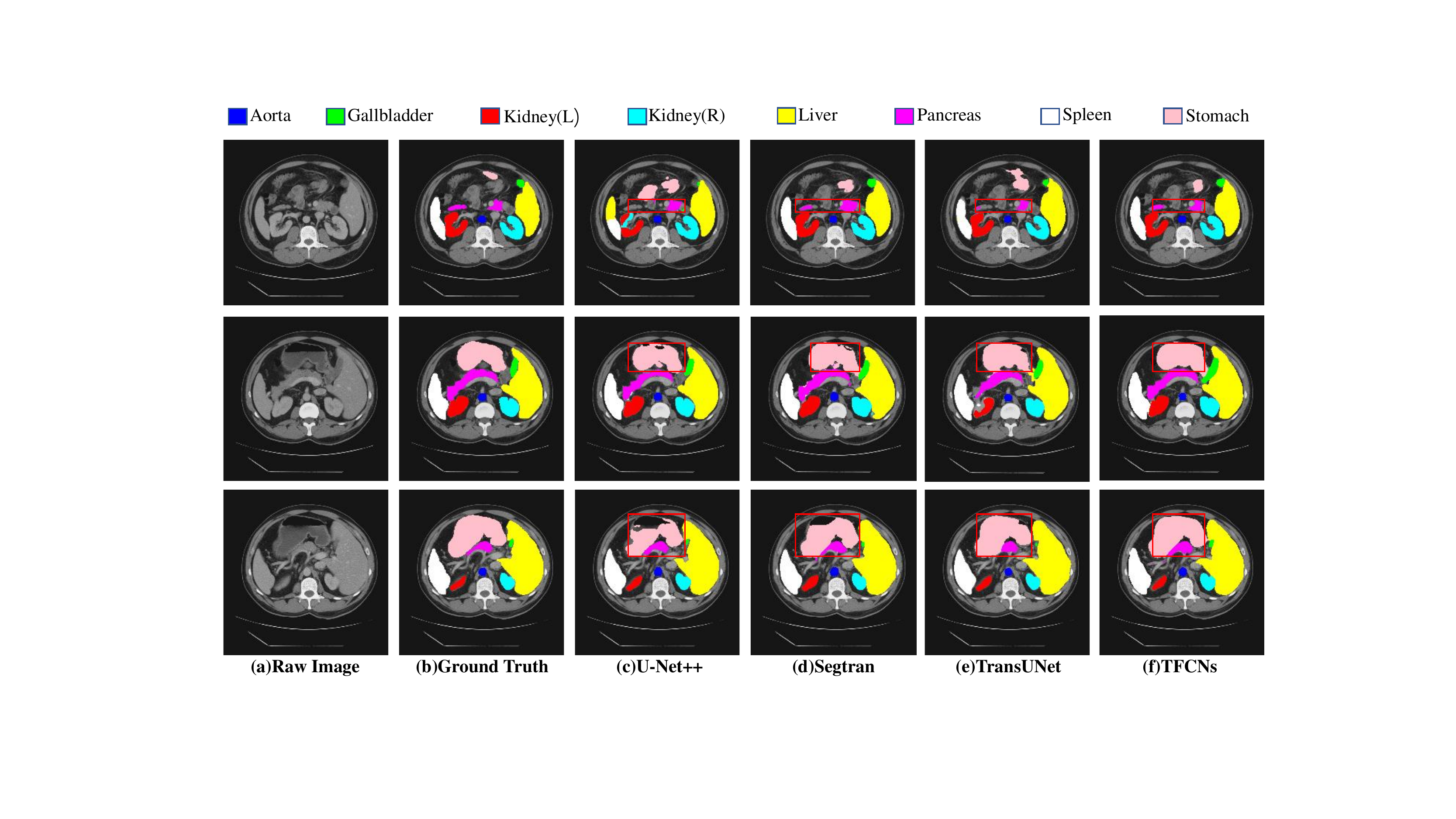}}
\caption{Qualitative results on Synapse Dataset. All columns respectively represent: (a) Raw Image; (b) Ground Truth; (c) U-Net++ \cite{zhou2018unet++}; (d) Segtran \cite{zheng2021rethinking}; (e) TransUNet \cite{chen2021transunet}; (f) TFCNs. Top row describes corresponding color of each organ in raw image.}
\label{fig3}
\end{figure}
%\noindent\textbf{Results and Visualization}\par
\noindent\textbf{Analysis of segmentation for COVID-19 infected areas}\par

\begin{figure}[H]
\centerline{\includegraphics[width=\columnwidth]{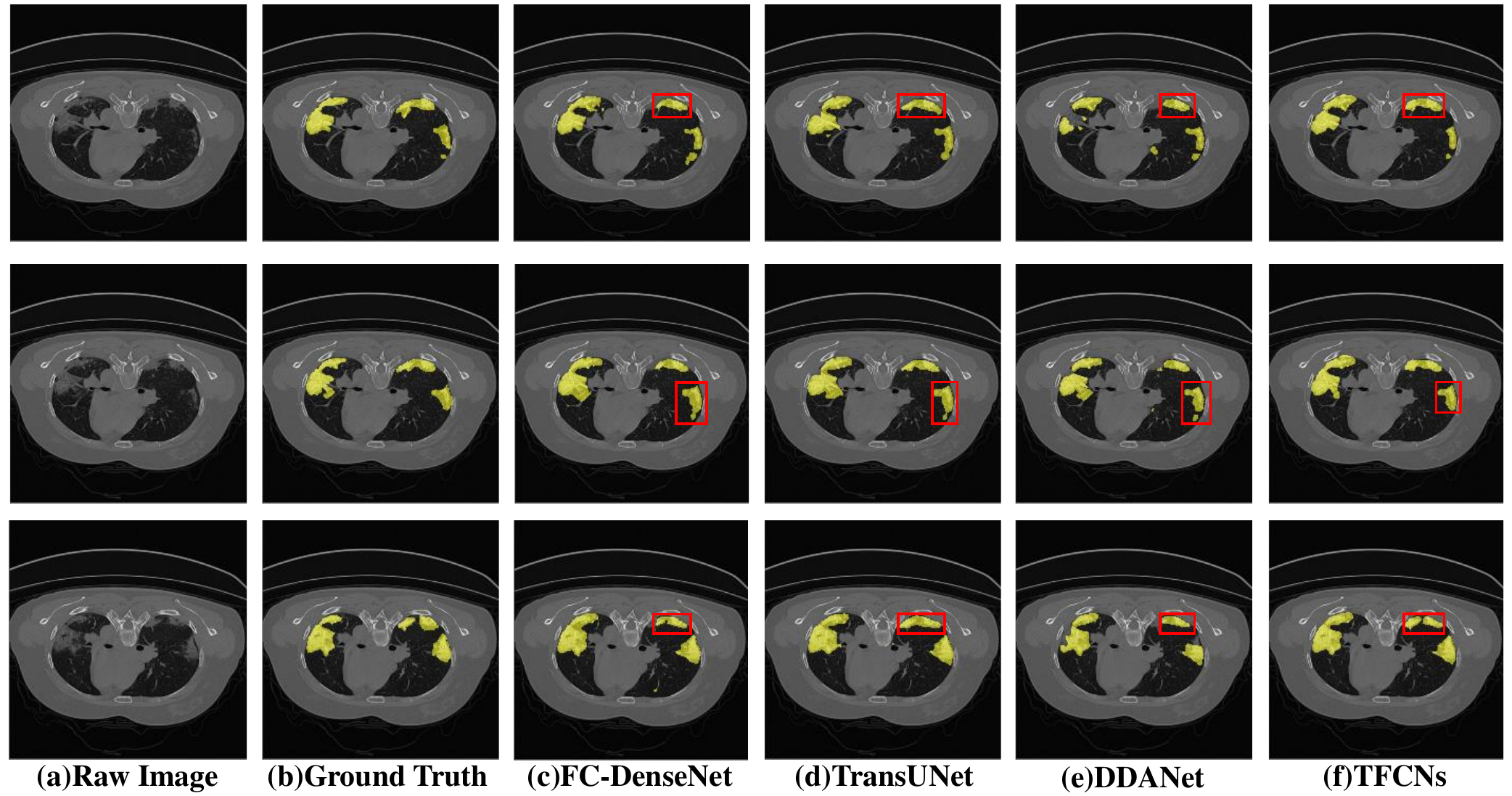}}
\vspace{-2mm}
\caption{Qualitative results on COVID-19 dataset. All columns respectively represent: (a) Raw Image; (b) Ground Truth; (c) FC-DenseNet \cite{zhang2018automatic}; (d) TransUNet \cite{chen2021transunet}; (e) DDANet \cite{tomar2020ddanet}; (f) TFCNs. Infected area in lung are labeled in yellow.}
\label{fig5}
\end{figure}

The performance of our model at the fine-grained feature is explored using the second dataset, i.e. the COVID-19 dataset. Since lesion features normally are more refined and scattered than organs which cover a large proportion of the image. Table \ref{table2} indicates that the capability of our method on the fine-grained target also reaches the SOTA level.
\vspace{-5mm}
\begin{table}[H]\centering
\caption{Performance comparison between our method and other state-of-the-art methods on  Covid-19 dataset (Dice score\% and Hausdorff distance in mm and Jaccard score\%). Avg means result average of all testing cases.}
\vspace{-1mm}
\label{table2}
\resizebox{0.65\columnwidth}{!}{
\begin{tabular}{c|c|c|c}
\hline
    Method      & Dice(Avg)$\uparrow$      & Hd95(Avg)$\downarrow$     & Jaccard(Avg)$\uparrow$   \\ \hline
    Segtran \cite{zheng2021rethinking}     & 75.35          & 41.18          & 60.35          \\
    Res-UNet \cite{xiao2018weighted}     & 73.53          & 47.54          & 59.86          \\
    DDANet \cite{tomar2020ddanet}      & 75.52          & 39.36          & 60.52          \\
    U-Net \cite{ronneberger2015u}        & 73.96          & 45.19          & 59.99          \\
    FC-DenseNet \cite{zhang2018automatic} & 71.13          & 54.72          & 55.52          \\
    AttU-net \cite{oktay2018attention}   & 74.70          & 48.36          & 60.26          \\
    TransUNet \cite{chen2021transunet}   & 72.19          & 52.51          & 57.22          \\
    TFCNs       & \textbf{75.55} & \textbf{37.32} & \textbf{60.74} \\ \hline
    \end{tabular}}
\end{table}
\vspace{-5mm}
Combining with the results visualized in Fig.\ref{fig5}, it can be seen that  DDANet \cite{tomar2020ddanet}, which is very close to our method in Dice coefficient and Jaccard coefficient, is very prone to under-segment, indicating that its segmented strategy may be choosing to ignore the pixels that are hard to distinguish. This strategy avoids the decrease inaccuracy caused by erroneous prediction, but it can easily cause false-negative that should be as least as possible in the diagnosis. TransUNet \cite{chen2021transunet} shows the over-segmentation in all rows. This may be the result of excessive utilization of contextual information in the local area which is suitable for continuous and regular objects such as organs, but these pieces of information are still too coarse for the lesion features. However, we solve this problem by adding the CLAB module.
\vspace{-5mm}

\subsection{Interpretability Analysis}
\label{sec4.3}

\begin{figure}[H]
\centerline{\includegraphics[width=\columnwidth]{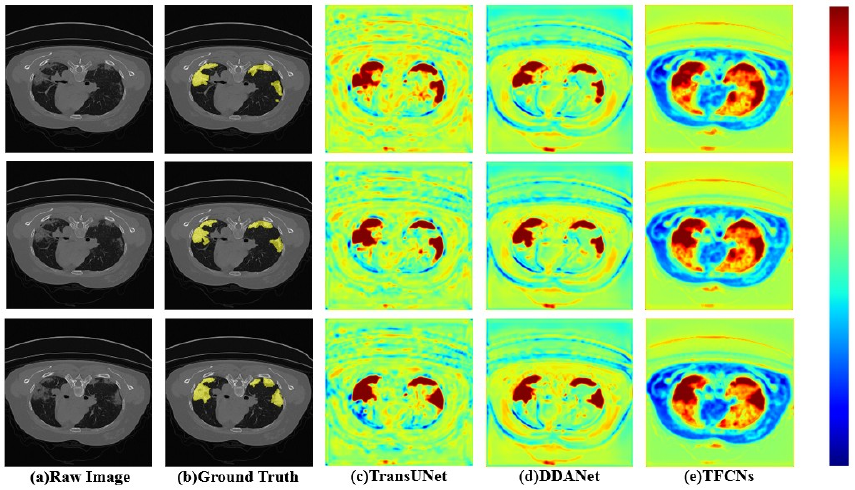}}
\caption{Heat map for interpretability analysis of different approaches on COVID-19 dataset. All columns respectively represents: (a) Raw Image; (b) Ground Truth; (c) TransUNet \cite{chen2021transunet}; (d) DDANet \cite{tomar2020ddanet}; (e) TFCNs. The colormap in right presents the degree of attention which increases from bottom to top. Infected area in lung are labeled using yellow in ground truth.}
\label{fig6}
\end{figure}

Interpretability analysis is conducted on the COVID-19 dataset using Class Activation Mapping (CAM) \cite{zhou2016learning} to explore what causes the model to make the decision on every pixel and which area of feature map the model pays the most attention to when it segments the infected area. As shown in Fig.\ref{fig6}, TFCNs focuses its attention on the lung area and pay least attention in the area surrounding the lung which means our model can act as an expert that focuses on the lung area at the beginning, and then pays more and more attention to the infected area gradually (in Fig.\ref{fig6}, the lung area is light red, and the infected area is dark red) when it predicts.

As for TransUNet \cite{chen2021transunet}, although it can also focus its attention on the infected area when predicting, it disperses other attention to the entire image instead of gathering it to the lungs. This reflects the importance of our CLAB module because it relocates the attention of the model to the key area.

Moreover, even though DDANet \cite{tomar2020ddanet} achieves good results, it ignores most areas in the image except the infected area, indicating that it does not have the concept of lungs and only makes some mechanical predictions based on ground truth which will make it very difficult to predict at the edge of the infected area, resulting in a large number of false-negatives.

\subsection{Ablation Studies}
\label{sec4.4}
\noindent\textbf{Effectness of ResMLP}\par
It can be seen from Table \ref{table4} that the Dice coefficient increases by 1.87\% after using ResMLP. In addition, the Hausdorff distance drops by 8.38mm, indicating that ResMLP makes the semantic features in the Transformers propagate more complete, thereby increasing the contextual information extracted by the encoder, which promotes the performance of the overall structure.
\begin{table}[H]\centering
\caption{ Ablation study on verifying the effectness of ResMLP (Dice score\% and Hausdorff distance in mm and Jaccard score\%).}
\label{table4}
\begin{tabular}{c|c|c|c}
\hline
    & Dice$\uparrow$           & Hd95$\downarrow$           & Jaccard$\uparrow$        \\ \hline
ResMLP & \textbf{83.72} & \textbf{17.26} & \textbf{72.78} \\
MLP   & 81.85          & 25.64          & 70.57         \\ \hline
\end{tabular}
\end{table}

\noindent\textbf{Type of Attention Block}\par
As shown in Table \ref{table5}, the performance of the model is greatly improved after using the attention blocks (no matter what type it is), which means these attention blocks play a critical role in removing irrelevant and redundant information at skip connections. Moreover, it can be seen that after using the CLAB module we designed, the performance of the model is continuously improved, which demonstrates that our CLAB module can accurately locate the area containing more effective information in the feature map.
\begin{table}[H]\centering
\caption{Ablation study on different types of attention block in skip connection (Dice score\% and Hausdorff distance in mm and Jaccard score\%).}
\label{table5}
\begin{tabular}{c|c|c|c}
\hline
    Type of Attention Block & Dice$\uparrow$           & Hd95$\downarrow$           & Jaccard$\uparrow$        \\ \hline
    None                    & 80.16          & 28.40          & 68.53          \\
    CUAB\cite{wang2021cuab}                    & 81.83          & 25.68          & 70.72          \\
    CLAB                    & \textbf{83.72} & \textbf{17.26} & \textbf{72.78} \\ \hline
    \end{tabular}
\end{table}

\section{Conclusion}
\label{sec5}
To improve the performance of medical image segmentation, in this paper, we propose TFCNs based on Transformer \cite{vaswani2017attention} and FC-DenseNet \cite{zhang2018automatic}. And the internal structure of Transformers is modified by introducing residual connections to form RL-Transformer. This change can help enhance the receptive field and improve the coding ability of the model. In addition, a common module—CLAB, which is mainly composed of global average pooling and linear mapping, is designed in the network to filter out non-semantic features. Experimental results show that TFCNs which is the best among the baselines achieves a score of 83.72\% on the Dice coefficient and a score of 72.78\% on the Jaccard coefficient in terms of the Synapse dataset. The experiments are also conducted on the COVID-19 public dataset, and results show that TFCNs also has the state-of-the-art performance.

\section{Acknowledge}
\label{sec6}
This work was supported in part by the Natural Science Foundation of Fujian Province of China (No. 2020J01006), the National Natural Science Foundation of China (No. 61502402), and the Open Project Program of State Key Laboratory of Virtual Reality Technology and Systems, Beihang University (No.VRLAB2022\\AC04).

\bigskip
\bibliographystyle{splncs04}

\end{document}